\documentclass[10pt, conference]{IEEEtran}
\usepackage{amsmath}
\usepackage{amssymb}
\usepackage{amsfonts}
\usepackage{amsthm}
\usepackage{epsfig}
\usepackage{subfigure}
\usepackage{calc}
\usepackage{color}
\usepackage[all]{xy}
\usepackage{bm}
\usepackage{algorithmicx}
\usepackage{algpseudocode}
\usepackage{algorithm}
\usepackage{enumerate}
\usepackage{cite}
\usepackage{multicol}
\usepackage{graphicx}
\usepackage{stfloats}
\usepackage{bbm}
\usepackage{enumitem}
\newlist{Properties}{enumerate}{2}
\setlist[Properties]{label=Property \arabic*.,itemindent=*}

\newtheorem{defn}{Definition}
\newtheorem{thm}{Theorem}[section]
\newtheorem{cor}[thm]{Corollary}
\newtheorem{prop}{Proposition}

\newtheorem{lem}[thm]{Lemma}
\newtheorem{conj}[thm]{Conjecture}
\newtheorem{constr}[thm]{Construction}
\newtheorem{note}{Remark}
\newtheorem{example}{Example}
\newcommand{\bit}{\begin{itemize}}
\newcommand{\eit}{\end{itemize}}
\newcommand{\bcor}{\begin{cor}}
\newcommand{\ecor}{\end{cor}}
\newcommand{\beq}{\begin{equation}}
\newcommand{\eeq}{\end{equation}}
\newcommand{\beqn}{\begin{equation*}}
\newcommand{\eeqn}{\end{equation*}}
\newcommand{\bea}{\begin{eqnarray}}
\newcommand{\eea}{\end{eqnarray}}
\newcommand{\bean}{\begin{eqnarray*}}
\newcommand{\eean}{\end{eqnarray*}}
\newcommand{\ben}{\begin{enumerate}}
\newcommand{\een}{\end{enumerate}}
\newcommand{\bdefn}{\begin{defn}}
\newcommand{\edefn}{\end{defn}}
\newcommand{\bnote}{\begin{note}}
\newcommand{\enote}{\end{note}}
\newcommand{\bprop}{\begin{prop}}
\newcommand{\eprop}{\end{prop}}
\newcommand{\blem}{\begin{lem}}
\newcommand{\elem}{\end{lem}}
\newcommand{\bthm}{\begin{thm}}
\newcommand{\ethm}{\end{thm}}
\newcommand{\bconj}{\begin{conj}}
\newcommand{\econj}{\end{conj}}
\newcommand{\bconstr}{\begin{constr}}
\newcommand{\econstr}{\end{constr}}
\newcommand{\bpf}{\begin{proof}}
\newcommand{\epf}{\end{proof}}

\title{On the Polarizing Behavior and Scaling Exponent of Polar Codes with Product Kernels}
\author{
\IEEEauthorblockN{Manan Bhandari, Ishan Bansal and V. Lalitha\\}
\IEEEauthorblockA{SPCRC, International Institute of Information Technology Hyderabad\\
Email: \{manan.bhandari@research.iiit.ac.in, ishan.bansal@students.iiit.ac.in, lalitha.v@iiit.ac.in\}\\}
\vspace{-0.7cm}
}
\date{\today}

\begin{document}
\maketitle

\begin{abstract}
Polar codes, introduced by Arikan, achieve the capacity of arbitrary binary-input discrete memoryless channel $W$ under successive cancellation decoding. Any such channel having capacity $I(W)$ and for any coding scheme allowing transmission at rate $R$, scaling exponent is a parameter which characterizes how fast gap to capacity decreases as a function of code length $N$ for a fixed probability of error. The relation between them is given by $N\geqslant \alpha/(I(W)-R)^\mu$. Scaling exponent for kernels of small size up to $L=8$ have been exhaustively found. In this paper, we consider product kernels $T_{L}$ obtained by taking Kronecker product of component kernels. We derive the properties of polarizing product kernels relating to number of product kernels, self duality and partial distances in terms of the respective properties of the smaller component kernels. Subsequently, polarization behavior of component kernel $T_{l}$ is used to calculate scaling exponent of $T_{L}=T_{2}\otimes T_{l}$. Using this method, we show that $\mu(T_{2}\otimes T_{5})=3.942.$ Further, we employ a heuristic approach to construct good kernel of $L=14$ from kernel having size $l=8$ having best $\mu$ and find $\mu(T_{2}\otimes T_{7})=3.485.$    
\end{abstract}

\section{Introduction}
Polar codes, introduced in \cite{arikan2009channel} are a family of codes which achieve capacity of binary input memory-less symmetric (BMS) channels with low complexity encoding and decoding algorithms. Polar codes are constructed based on polar transform given by kernel $T_{2}$ =
$\begin{bmatrix}
1 & 0 \\
1 & 1
\end{bmatrix}$ and its Kronecker product taken $n$ times $T_{2}^{\otimes n} = G_{N}$. The polar transform given by $G_N$ transforms a set of $N$ independent copies of the BMS channel $W$ into $N$ bit channels which are either noiseless or full-noisy. The fraction of the bit channels which are noiseless approaches the symmetric capacity $I(W)$ as $N \rightarrow \infty$. 
In order to prove that polar codes achieve capacity, it is enough to consider successive cancellation decoder. 
However, in practice, successive cancellation list decoder is employed for better error performance \cite{tal2011list}. The phenomenon of channel polarization holds for any kernel $T_{l}$ of size $l\times l$ under certain conditions on the kernel. Therefore, any polar code of length of the form $N = l^{n}$ can be constructed.

\subsection{Scaling Exponent}
To analyze the performance of polar codes, the parameters of interest are: rate $R$, block length $N$, and block error probability $P_{e}$. For fixed $W$ and $R$, error exponent $\gamma$ characterizes how fast $P_{e}$ converges to 0 as a function of $N$.
Error exponents for polar codes obtained from a kernel have been derived in \cite{korada2010polar}.

For fixed $W$ and $P_{e}$, scaling exponent characterizes how fast the rate of the polar code $R$ can approach the capacity as a function of block length $N$. The following inequalities give the relation between $N$ and $R < I(W)$ in terms of the scaling exponent

\begin{equation}
    \frac{\alpha_{1}}{(I(W)-R)^{\mu_{1}}}\leqslant N\leqslant \frac{\alpha_{2}}{(I(W)-R)^{\mu_{2}}}
\end{equation}
where $\alpha_{1}$ and $\alpha_{2}$ are positive constants depending on $P_{e}$ and $I(W)$. 

It is known from \cite{mondelli2016unified} that scaling exponent for random codes equals 2 and shown in \cite{fazeli2017binary} that $\mu$ for polar codes approaches 2 as $l\to \infty$ with high probability for BEC channel over random choice of the kernel. It is already known that $\mu = 3.627$ for conventional polar codes (Arikans $T_2$ kernel) over BEC \cite{fazeli2017binary}. Recently, a class of self-dual binary kernels were introduced in \cite{yao2019explicit} in which large kernels of size $2^{p}$ were constructed with low $\mu$ and it was shown that $\mu=3.122$ for $L=32$ and $\mu\simeq 2.87$ for $L=64$.  

\subsection{Polar Codes with Product Kernels}
Calculating scaling exponent of large kernels in general is a NP hard problem \cite{fazeli2014scaling}. In order to obtain polarization behavior and therefore the scaling exponent efficiently, a class of kernels is considered, which we term as product kernels. This kernel is formed by taking kronecker product of smaller binary kernels termed as component kernels. We define product kernels as follows
\vspace{-0.01in}
\begin{equation}T_{L} = T_{l_{1}} \otimes T_{l_{2}} \dotsm \otimes T_{l_{m}}
\end{equation}
where $T_{l_{i}}$ is the component kernel of size $l_{i}\times l_{i}, i=1,\dotsm, m$. The dimension of the product kernel is $L\times L$ where $L=l_{1}\cdot l_{2}\dotsm l_{m}$ and length of the polar code is $N=L^{n}$. 
There exists a related and much general class of polar codes known as multi-kernel polar codes \cite{benammar2017multi}, where the transformation matrix $G_N$ itself is formed by taking tensor product of kernels with different lengths.

\subsection{Our Contributions}
In this paper, we find the number of polarizing product kernels for a given $L=l_{1}\cdot l_{2}\dotsm l_{m}$ given the number of polarizing component kernels. We also prove the self duality of these kernels given the self-dual property of component kernels (self-dual kernels are explained later in the paper). We find partial distance and polarization behavior of product kernels of the form $T_{L}=T_{2}\otimes T_{l}$ in terms of partial distance and polarization behavior of $T_{l}$ respectively. 

Further, scaling exponent is calculated using a recursive function of polarization behavior which gives us $\mu = 3.942$ for $L=10$. We then propose a heuristic approach to find good kernel of the size $L=14$ from the kernel of size $l=8$ having the best scaling exponent. A $14\times 14$ kernel having $\mu=3.485$ is constructed by this method. We have also analyzed and plotted the variation of scaling exponent as kernel size increases.

\section{Polarizing Product Kernels}
A polarizing kernel is an $l\times l$ binary matrix which is non-singular and not upper triangular under any column permutation \cite{korada2010polar}. For any $l\times l$ kernel, number of polarizing kernels is
\begin{equation}
 M_{T_{l}} = 2^{\frac{l(l-1)}{2}}(\prod_{i=1}^l(2^i-1)-l!)
\end{equation}
where the first term is number of non-singular matrices and second term is number of upper triangular matrices \cite{fazeli2014scaling}.

In this section, we will derive the number of polarizing product kernels for polar codes in terms of the number of polarizing kernels for any $l\times l$ kernel, $M_{T_l}$.

\begin{prop}For a product kernel $T_{L}$ to be upper triangular, each of the component kernels $T_{l_{i}}$ have to be upper triangular. \end{prop}

\bpf Let $T_{L}=T_{l_{1}}\otimes T_{l_{2}}$. After taking the Kronecker product, $T_{L}$ consists of multiple sub-matrices of $T_{l_{2}}$, depending on the values of $T_{l_{1}}$. For $T_{L}$ to be upper triangular, all the entries in the lower half have to be zero. Values of all the sub-matrices in the lower half, except the ones on the diagonal will be zero only when all the entries in the lower half of $T_{l_{1}}$ are zero. The values that remain are the lower-half values of sub-matrices lying on the diagonal, which should also be zero. This will happen only when $T_{l_{2}}$ should itself be upper-triangular. Hence, both $T_{l_{1}}$ and  $T_{l_{2}}$ should be upper triangular for $T_L$ to be an upper triangular matrix. Similarly, this proof can be extended for any general $T_{L} = T_{l_{1}} \otimes T_{l_{2}} \dotsm \otimes T_{l_{m}}.$ \epf

 \begin{prop}For a product kernel to be non-singular, each of the component kernels have to be non-singular.\end{prop}
 \bpf  For any two matrices A and B, rank($A\otimes B$) = rank(A).rank(B). All the component matrices are non-singular and full rank matrices. Hence, $T_{L}$ should be non-singular and of full rank.\epf
 Therefore, total number of polarizing product kernels for $T_{L}= T_{l_{1}} \otimes T_{l_{2}} \dotsm \otimes T_{l_{m}}$ are
\begin{equation}
    M_{T_{L}} = \prod_{j=1}^mM_{T_{l_{j}}} =  \prod_{j=1}^m \left (2^{\frac{l_{j}(l_{j}-1)}{2}}(\prod_{i=1}^{l_{j}}(2^i-1)-l_{j}!) \right )
\end{equation}

\section{Polarization Behavior of Polar Codes with Product kernels}
\label{secfour}
Scaling exponent $\mu$ is a parameter which characterizes how fast gap to capacity decreases as a function of code length $N$ for a fixed probability of error. It is dependent on the polarization behavior of the polar codes. For any kernel $T_{l}$, channel $W$ (with erasure probability $z$) polarizes into $l$ bit channels $W_{1},W_{2},\dotsm, W_{l}$. The erasure probabilities of each of these bit channels is given by ${p_{1}(z),p_{2}(z),\dotsm,p_{l}(z)}$ respectively and this set is known as polarization behavior of $T_{l}$ which determines the scaling exponent of that kernel. 

\begin{figure}[!t]
\centering
\includegraphics[scale=0.45]{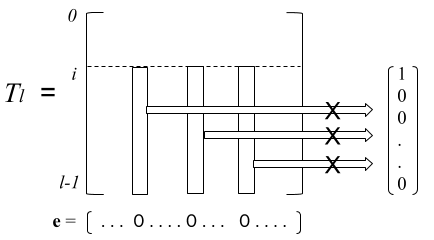}
\caption{Illustration of {\bf e} killing the channel $W^i$}
\label{fig:my_label1}
\end{figure}

\begin{defn}[Erasure Pattern]
    An erasure pattern {\bf e} is a binary vector of length $l$. If $e_{i}=1$, the $i^{th}$ copy of $W$ is erased. The number of erasures in vector {\bf e} is defined as weight of {\bf e} and denoted by $wt({\bf e})$ and number of non-erasures is $l-wt({\bf e})$. Hence, probability of any erasure pattern {\bf e} is $z^{wt({\bf e})}(1-z)^{l-wt({\bf e})}$.
\end{defn}
\begin{defn}  
    An erasure pattern {\bf e} is said to kill the bit channel $W^i$ if there is no linear combination of non-erased columns in $T_{l}[i:l-1]$ that gives the vector $[1,0,0 \dotsm, 0]^t$ which is of length $(l-i)$ where $T_{l}[i:l-1]$ is sub-matrix of $T_{l}$ with rows only from $i$ to $l-1$ as shown in Fig  \ref{fig:my_label1}. Let us denote $[1,0,0 \dotsm, 0]^t$ of length $(l-i)$ as $Y_{l-i}$ for further use in the paper. 
\end{defn}
    
Number of such erasure patterns of $wt({\bf e})=w$ that satisfy the above killing condition is denoted by $E_{i,w}$. The erasure probability of corresponding bit channel $W^i$ is given by

\vspace{-0.15in}
\begin{equation}
\label{equone}
    p_i(z) = \sum_{w=0}^{l}E_{i,w}z^w(1-z)^{l-w}
\end{equation}
\vspace{-0.1in}

A straightforward way to find $p_{i}(z)$ of any product kernel $T_L = T_{l_1} \otimes T_{l_2}$ is to use the composite function property of polarization behavior. Let the polarization behavior of $T_{l_1}$ be $f_{j}(z)$ and of $T_{l_2}$ be $g_{k}(z)$, then polarization behavior of $T_L$ is $p_{i}(z)=f_{j}(g_{k}(z))$. Another way to find $p_{i}(z)$ is to calculate all $E_{i,w}$ for all weights of the erasures of $T_{L}$ where $0\leqslant i<L$. The computational complexity of doing so is NP-hard in general. In this paper, we propose an alternate approach which finds the polarization behavior of $T_{L}$ based on the $E_{i,w}$ of the component kernels. We can infer some properties of the product kernel based on $E_{i,w}'s$ which will be discussed later. We find an analytical method to calculate the number of erasure patterns $E_{i,w}^{T_{L}}$ in terms of $E_{i,w}'s$ of its component kernels.

For further calculation, we assume $T_{L} = T_{2} \otimes T_{l}; \hspace{2mm} L=2l$.
Let the erasure pattern ${\bf e}_{T_{L}} = [{\bf e}_{l} \hspace{1mm} {\bf e}_{l^{'}}]$  where ${\bf e}_{l}$ is the erasure pattern with length $l$ and weight $w_{1}$ and ${\bf e}_{l^{'}}$ is the erasure pattern with length $l$ and weight $w_{2}$. Hence the weight of ${\bf e}_{T_{L}}$ is $w = w_{1}+w_{2}$. As we know, $T_{2}$ and $T_{l}$ need to satisfy polarizing conditions, there exists only one valid $T_{2}$ for which calculations of $E_{i,w}^{T_{L}}$ are to be done. 

Note : There can be 4 sub-cases in each case:
\begin{itemize}
\item ${\bf e}_{l}$ kills but ${\bf e}_{l^{'}}$ doesn't kill channel $W^{i}$
\item ${\bf e}_{l}$ doesn't kill but ${\bf e}_{l^{'}}$ kills channel $W^{i}$
\item Neither ${\bf e}_{l}$ or ${\bf e}_{l^{'}}$ kill channel $W^{i}$ 
\item Both ${\bf e}_{l}$ and ${\bf e}_{l^{'}}$ kill channel $W^{i}$
\end{itemize}

It can be noted that if either ${\bf e}_{l}$ or ${\bf e}_{l^{'}}$ don't kill $W^{i}$ implies that there exists a linear combination of non-erased columns which gives $Y_{l-i}$. This implies that ${\bf e}_{T_{L}}$ will also not kill the channel in these cases leaving only one case in which both ${\bf e}_{l}$ and ${\bf e}_{l^{'}}$ kill the channel. We find polarization behavior of $T_{L}$ in terms of $T_{l}$. 

%\begin{enumerate}
$T_{2} = \begin{bmatrix} 1 & 0 \\1 & 1 \end{bmatrix}$ \hspace{4mm}
$T_{L} = \left[\begin{array}{c|c} T_{l} & 0\\ \hline T_{l} & T_{l} \end{array} \right]$.

\noindent \textbf{1) $0\leqslant i\leqslant l-1$ (upper half) :} \\
   \noindent To find $E_{i,w}$ in this case, we first find the number of erasures of weight $w$ for which at least one of the linear combination of non-erased columns give $Y_{2l-i}$ and then subtract it from total number of combinations possible to get $E_{i,w}$. If $wt({\bf e}_{T_{L}}) = 2l-1$, then selecting only one column corresponding to ${\bf e}_{l}$ or ${\bf e}_{l'}$ cannot result in a null vector for rows $l$ to $2l-1$. Therefore, $E_{i,2l-1}=0$. If $wt({\bf e}_{T_{L}}) < 2l-1$, the number of ${\bf e}_l$ such that $wt({\bf e}_l) = l-j$ and which result in $Y_{l-i}$ from rows $0$ to $l-1$ by taking sum of all $j$ non-erased columns is given by  
    \begin{equation*}
        X_{i,j} = \left(\binom{l}{j}- (E_{i,l-j})\right)-\left(\binom{l}{j-1}- (E_{i,l-j+1})\right) .
    \end{equation*}
    It is easy to see that $X_{i,j}\geqslant 0$. Now to construct an erasure pattern of weight $w$ which kills channel $W^i$, an erasure pattern ${\bf e}_{l}$ as above is considered. If $j$ non-erased columns are picked from ${\bf e}_{l}$ whose sum results in $Y_{l-i}$ for the upper half, then the same $j$ non-erased columns have to be picked from ${\bf e}_{l'}$ to obtain null vector for rows from $l$ to $2l-1$. Therefore, from the remaining $(2l-2j)$ columns, select $(2l-w-2j)$ non-erased columns to result in an erasure pattern of weight $w$ which kills channel $i$.
    Hence, the number of erasures of weight $w$ killing the channel $W^i$ are given by 
    \begin{equation}\label{eqnuse1}
        E_{i,w} = \binom{2l}{2l-w} - \sum_{j=1}^{(2l-w)/2}X_{i,j}\cdot \binom{2l-2j}{2l-w-2j}.
    \end{equation}

    \begin{example}
     To find $E_{i,2}$ value for $T_{10}$ :\\
     The required number of non-erased columns are 8. Let ${\bf e}_{l} = [1\ 0\ 0\ 1\ 0]$ be a combination giving $Y_{l-i}$ in $T_{5}$. In order to make a null vector from rows $l$ to $2l-1$ in $T_{10}$, we need to choose columns $(7,8,10)$ as non-erased columns corresponding to the above columns of $(2,3,5)$. Now, $6$ non-erased columns out of $8$ are already chosen. Others can be selected arbitrarily in $\binom{10-6}{10-6-2} =6$ ways. Similar procedure is repeated for all erasures of $T_5$ to get the count of erasures which don't kill channel $W^{i}$. 
    \end{example}
    
\noindent  \textbf{2) $l\leqslant i\leqslant 2l-1$ (lower half) :}\\
    For this case, note that for an erasure of weight $w_{0}$ killing the channel in $T_{l}$, each non-erased column can be either a part of $T_{l}$ corresponding to ${\bf e}_{l}$ or $T_{l}$ corresponding to ${\bf e}_{l'}$. In short, each of these columns can be split between $T_{l}$ corresponding to ${\bf e}_{l}$ and ${\bf e}_{l'}$ with total weight still being $w_{0}$. Also, in $T_{L}$, any erasure of weight $w < w_{0}$ killing the channel can be made by repeating some of the non-erased columns in ${\bf e}_{l}$ and ${\bf e}_{l'}$. Number of ways of choosing these repeating columns is therefore $\binom{l-w_{0}}{2l-w-(l-w_{0})}$. Hence, the number of ways of choosing remaining non-repeating non-erased columns is $2^{2(l-w_{0})-(2l-w)}$. Therefore,
    \begin{equation}\label{eqnuse2}
        E_{i+l,w} = \sum_{w_{0}=max(2l-w,l)}^{2l}\binom{l-w_{0}}{l-w+w_{0}}\cdot 2^{w-2w_{0}}\cdot E_{i,w_{0}}.
    \end{equation}
      
    \begin{figure}[!ht]
    \centering
    \includegraphics[scale=0.35]{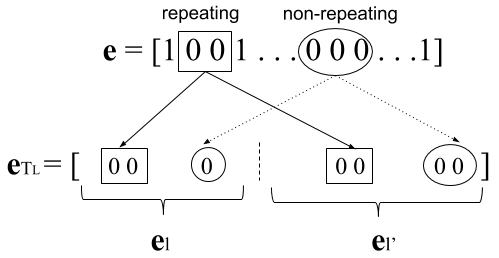}
    \caption{Representation of ${\bf e}_{T_L}$ formed by repeating and non-repeating non-erasures in ${\bf e}$.}
    \centering
    \label{fig:my_label_3}
    \end{figure}

    \begin{example}
    Let $l=5$ and ${\bf e}_{l}=[1\ 0 \ 0 \ 1\ 0]$ is a combination not giving $Y_{l-i}$ in $T_{5}$. In order to find $E_{i,5}$ for $T_{10}$, we note that the following erasure patterns of weight 5 are obtained based on ${\bf e}_l$ as follows:
    \begin{itemize}
       \item If columns $(2,3)$ are repeated, the corresponding columns are $(7,8)$. The non-repeating column can be either $5$ or $10$ giving $(2,3,5,7,8)$ and $(2,3,7,8,10)$
        \item If columns $(2,5)$ are repeated, we get $(2,3,5,7,10)$ and $(2,5,7,8,10)$
        \item Similarly, if columns $(3,5)$ are repeated, we get $(2,3,5,8,10)$ and $(3,5,7,8,10)$
    \end{itemize} 
    In total we get $\binom{3}{2}\cdot 2^{1} = 6$ combinations. Similar procedure is repeated for all erasure patterns of weight $w \leq w_0$ to get $E_{i,w}$.
    \end{example}
%    \end{enumerate}
%\end{enumerate}

\section{Self-Duality and Partial Distance of Product Kernels}
Self-dual kernels are a special class of polarizing kernels having symmetric polarization behaviors. In this section, we prove that product of self-dual kernels is also self dual. We also characterize the partial distances of product kernels of the form $T_2 \otimes T_l$. In the following, we denote the span of $k$ vectors $v_1, v_2, \ldots v_k \in \mathbb{F}_2^l$ by $<v_1, v_2, \ldots, v_k>$. Also, dot product of $v_m$ and $v_n$ is denoted by $v_m. v_n$ and defined as $v_m.v_n = \sum_{p = 1}^l v_{m,p} v_{n,p}$.

\begin{defn} [Self-Dual Kernel]
Let us denote any $l\times l$ kernel $T_l = [g_{1}^T, g_{2}^T, \dotsm, g_{l}^T]^T$. Kernel codes $C_i$ are defined as $C_i = \langle g_{i+1},g_{i+2}, \dotsm,g_l \rangle$ for $0\leqslant i<l$ and $C_l = \{0\}$. This kernel is said to be self-dual if $C_i = C_{l-i}^\perp$ for all $0\leqslant i\leqslant l$. 
\end{defn} 
Some of the properties of self dual kernel proved in \cite{yao2019explicit} are:
\begin{Properties}
    \item $\forall_w : E_{i,w} + E_{l+1-i,l-w}\leqslant \binom{l}{w} $ for $i=1,\dotsm,l$.
    \item $f_{l+1-i}(z) = 1 - f_{i}(1-z)$ for $i=1,\dotsm,l$.
    \item By constructing just one half of the kernel, other half can be obtained by the symmetric polarization behavior stated in the duality theorem (property 2). When $z$ is close to 0, $f_i(z)$ is dominated by the partial distance (defined in later section). Therefore the aim is to construct rows of lower half to maximize the partial distance to make $f_i(z)$ polarize towards 0. 
\end{Properties}

As we know that if $T_l$ is self-dual, $C_i = C_{l-i}^\perp$, we have the following: 
\begin{itemize}
    \item \textbf{P1} : For any general $i$,  $C_i = \langle g_{i+1},\dotsm,g_l \rangle$ and $C_{l-i} = \langle g_{l-i+1}, \dotsm,g_l \rangle$. If $C_i = C_{l-i}^\perp$, it follows that \\ $g_j \cdot g_k = 0$ for $j \in [i+1,l]$ and $ k \in [l-i+1,l]$ and $g_j \cdot g_k \neq0$ for $j \in [1,\frac{l}{2}]$ and $ k \in [i+1,l-i]$.
    \item \textbf{P2} : We know that $dim(C_i) = l- dim(C_{l-i})$.
\end{itemize}

\begin{thm}
If $T_{l_1} $ and $T_{l_2}$ are self-dual kernels, then $T_{l_1} \otimes T_{l_2}$ is also a self-dual kernel.
\end{thm}
\bpf Let $T_L = T_{l_1} \otimes T_{l_2}$ be the product kernel where $L=l_1l_2$. $T_{l_1}= [f_{1}^T, \dotsm, f_{l}^T]^T$ and $T_{l_2}= [g_{1}^T, \dotsm, g_{l}^T]^T$ are self-dual kernels satisfying all its properties individually. 

 \begin{center}
$T_{L} = \begin{bmatrix} 
f_{1,1}g_1 & f_{1,2}g_1 & \dotsm & f_{1,l_{1}}g_1\\
f_{1,1}g_2 & f_{1,2}g_2 & \dotsm & f_{1,l_{1}}g_2\\
&\vdots\\
f_{1,1}g_{l_{2}} & f_{1,2}g_{l_{2}} & \dotsm & f_{1,l_{1}}g_{l_{2}}\\
 &\vdots\\
f_{{l_1/2},1}g_1 & f_{{l_1/2},2}g_1 & \dotsm & f_{{l_1/2},l_{1}}g_1\\
&\vdots\\
f_{{l_1/2},1}g_{l_{2}} & f_{{l_1/2},2}g_{l_{2}} & \dotsm & f_{{l_1/2},l_{1}}g_{l_{2}}\\
&\vdots\\
f_{l_1,1}g_1 & f_{l_1,2}g_1 & \dotsm & f_{l_1,l_{1}}g_1\\
&\vdots\\
f_{l_1,1}g_{l_{2}} & f_{l_1,2}g_{l_{2}} & \dotsm & f_{l_1,l_{1}}g_{l_{2}}\\
\end{bmatrix}$.
\end{center}

We divide the proof into two cases:

\noindent {\bf Case 1:} $i = s l_2, 0 \leq s \leq \frac{l_1}{2}$. In this case, we have 
$ C_{i}  =  \langle (f_{s+1}\otimes g_1), (f_{s+1}\otimes g_2), \dotsm (f_{s+1} \otimes g_{l_2}), (f_{s+2}\otimes g_1), \dotsm  (f_{s+2}\otimes g_{l_2}),
\dotsm  (f_{l_1}\otimes g_1), \dotsm (f_{l_1}\otimes g_{l_2})\rangle$ and $C_{l_1l_2 - i} = \langle (f_{l_1-s+1}\otimes g_1), (f_{l_1-s+1}\otimes g_2), \dotsm (f_{l_1-s+1} \otimes g_{l_2}),
(f_{l_1-s+2}\otimes g_1), \dotsm  (f_{l_1-s+2}\otimes g_{l_2}), \dotsm  (f_{l_1}\otimes g_1), \dotsm (f_{l_1}\otimes g_{l_2})\rangle$.

Dot product of a vector in $C_i$ and $C_{l_1l_2 - i}$ is given by $(f_u \otimes g_m). (f_v \otimes g_n) = (\sum_{p=1}^{l_2}  g_{m, p} g_{n,p})( \sum_{k=1}^{l_1}f_{u,k}f_{v,k})$ where $u \in [s+1,l_1], v \in [l_1-s+1,l_1]$ and $1\leqslant m,n\leqslant l_2$. Applying P1 to kernel $T_{l_1}$, the dot product evaluates to zero.

\vspace{0.2in}

\noindent {\bf Case 2:} $ i = s l_2 + t, 0 \leq s \leq \frac{l_1}{2}, 0 < t < l_1$. In this case, we have $C_{i} = \langle (f_{s+1} \otimes g_{t+1}),  \dotsm (f_{s+1} \otimes g_{l_2}), (f_{s+2}\otimes g_1), \dotsm  (f_{s+2}\otimes g_{l_2}), \dotsm  (f_{l_1}\otimes g_1), \dotsm (f_{l_1}\otimes g_{l_2})\rangle$ and
$C_{l_1l_2 - i} = \langle (f_{l_1-s}\otimes g_{l_2-t+1}),  \dotsm (f_{l_1-s} \otimes g_{l_2}), (f_{l_1-s+1}\otimes g_1), \dotsm  (f_{l_1-s+1}\otimes g_{l_2}), \dotsm  (f_{l_1}\otimes g_1), \dotsm (f_{l_1}\otimes g_{l_2})\rangle$.
The dot products of vectors in $C_i$ and $C_{l_1l_2 - i}$ fall in the following three categories: \\
(a) $(f_u \otimes g_m). (f_v \otimes g_n)$ where  $u \in [s+1,l_1], v \in [l_1-s+1,l_1]$. These dot products are zero by applying P1 to kernel $T_{l_1}$. \\
(b) $(f_u \otimes g_m). (f_v \otimes g_n)$ where  $u \in [s+2,l_1], v \in [l_1-s,l_1]$. These dot products are zero by applying P1 to kernel $T_{l_1}$.\\
 (c)  $(f_{s+1} \otimes g_m). (f_{l_1-s} \otimes g_n)$ where  $m \in [t+1,l_2], v \in [l_2-t+1,l_2]$. These dot products are zero by applying P1 to kernel $T_{l_2}$.

Based on the above arguments and P2, we can infer that $C_i = C_{l_1l_2-i}^\perp, i=0,1,\ldots, l_1l_2$. Hence, $T_{L}$ is a self-dual kernel.
\epf

\begin{defn}[Partial distance] For any $l\times l$ kernel $T_l$, $i^{\text{th}}$ partial distance is defined as $d_i = d_H(g_i,C_i)$ for $i=1, \dotsm ,l-1$ and $d_l=d_H(g_l,0)$ 
\end{defn}

As defined before, $C_i = \langle g_{i+1},g_{i+2}, \dotsm,g_l \rangle$ for $0\leqslant i<l$ and $C_l = \{0\}$. When $z$ is close to 0 in \eqref{equone}, the polynomial $p_i(z)$ is dominated by the first non-zero term $E_{i,w}z^w(1-z)^{(l-w)}.$ From \cite{yao2019explicit}, we know that the first non-zero coefficients of $p_i(z)$ is $E_{id_i}$. For construction of self-dual kernel, we aim to maximize the partial distance to make $p_i(z)$ polarize to 0.

It is clear from the definition of partial distance that $wt(g_i)\geqslant d_i$. This property will be used later in the proof of the below theorem.

\begin{thm}
If partial distances of the component kernel $T_l$ are $[d_1,d_2 \dotsm, d_l]$, then partial distances of $T_L = T_2 \otimes T_l$ are $[d_1, \dotsm, d_l, 2d_1 \dotsm 2d_l]$.
\end{thm}
\bpf Let $T_l = [g_{1}^T, \dotsm, g_{l}^T]^T$, $T_L = [G_{1}^T, \dotsm, G_{l}^T]^T$ and its partial distances be $[D_1,D_2 \dotsm, D_{2l}]$. We know that $T_{L} = \left[\begin{array}{c|c} T_{l} & 0\\ \hline T_{l} & T_{l} \end{array} \right]$. Let us prove by dividing it into two parts: \\
  {\bf 1) $0\leq i<l$:} There can be 3 sub-cases depending on how vectors are chosen from kernel codes $C_i$. We will find partial distance for each case. \\
   (a) When vectors only from row $(i+1)$ to $(l-1)$ (upper half) are chosen from the kernel codes : Let the linear combination of the vectors chosen be $v_1$ which will be of the form $(C_i,0)$. The partial distance of the left half will be same as the corresponding partial distance in $T_l$ and will be zero for right half.
        \begin{equation*}
            D_{i_1} = d_H(g_i,C_i) + d_H(0,0) \geqslant d_i.
        \end{equation*}
       (b) When vectors only from row $l$ to $(2l-1)$ (lower half) are chosen from the kernel codes : Let the linear combination of the vectors chosen be $v_1$. From triangle inequality,
        \begin{equation*}
        \begin{split}
            D_{i_2} & = d_H(g_i,v_1) + d_H(0,v_1) \\
            & = d_H(g_i,v_1) + d_H(v_1, 0)\geqslant d_H(g_i,0).
        \end{split}
        \end{equation*}
        It is clear that $d_H(g_i,0) \geqslant d_i$. Hence, $D_{i_2}\geqslant d_i$. \\
        (c) When vectors from both upper half and lower half are chosen from the kernel codes : Let the linear combination of the vectors chosen be $v_1$ and $v_2$ respectively.
        \begin{equation*}
        \begin{split}
            D_{i_3} & = d_H(g_i,v_1\oplus v_2) + d_H(0,v_2) \\
            & = d_H(g_i\oplus v_1, v_2) + d_H(v_2,0).
        \end{split}
        \end{equation*}
        Using triangle inequality and definition of $d_i$, we can say that $d_H(g_i\oplus v_1,v_2) + d_H(v_2,0)\geqslant d_H(g_i\oplus v_1, 0)$ and $d_H(g_i\oplus v_1, 0) = d_i$. Hence, $D_{i_3}\geqslant d_i$.
    From all these sub-cases, we can conclude that  min$(D_{i_1},D_{i_2},D_{i_3}) = d_i$.\\
   {\bf 2) $l< i\leq 2l$ :} In the lower half, the left half and right half are same. It is straightforward that the minimum distance $D_i$ will be the sum of minimum distance in both the halves. For any row $i$, we already know the minimum distances separately for both halves. Hence $D_i = 2d_{i-l}$. 
\epf

\section{Scaling Exponent of Product Kernels}
In this section, we first quickly review the procedure given in \cite{hassani2014finite} for calculating the scaling exponent. Consider a BEC channel $W$ with erasure probability $z$. Let $Z_n$ denotes the random process corresponding to the evolution of the Bhattacharaya parameters and $f_n(z,a,b)$ denote the fraction of unpolarized channels with thresholds $a$ and $b$, i.e., $ f_{n}(z,a,b) = Pr(Z_{n}\in [a,b])$.
 The function satisfies the following recursion with $n$ in terms of the polarization behavior: 
\begin{equation*}
    f_{n+1}(z,a,b) = \frac{\sum_{i=0}^{l-1}f_{n}(p_{i}(z),a,b)}{l}
\end{equation*}
with $f_{0}(z,a,b) = \mathbbm{1}_{z\in[a,b]}$. Assuming that there exists $\mu \in (0, \infty)$ such that for any $z,a,b \in (0,1)$ with $a<b$, the following limit exists in $(0, \infty)$ 
\begin{equation*}
    f(z,a,b)=\lim_{n\to \infty}l^{n/\mu}f_{n}(z,a,b).
\end{equation*}
The scaling exponent can be computed by solving the following equation numerically with appropriate initialization and stopping criterion:
\begin{align}
    \label{eqnuse3} l^{\frac{-1}{\mu}}f(z,a,b) = \frac{\sum_{i=0}^{l-1}f(p_{i}(z),a,b)}{l}.
\end{align}

The recursive function $f(z,a,b)$ is obtained by iterating through the procedure until the stopping condition $\|f_{n+1}(z)-f_{n}(z)\| \leqslant 10^{-8}$. It has been observed in \cite{hassani2014finite} that even for moderate values of $n$ ($8\leqslant n\leqslant10$), the function converges well\footnote{The stopping condition is $\|f_{n+1}(z)-f_{n}(z)\| \leqslant 10^{-10}$ in \cite{hassani2014finite}.}.

The method described in the Section \ref{secfour} to calculate polarization behavior is now used to calculate scaling exponent for any product kernel of the form $T_2 \otimes T_l$. We start with $L=10$, as best $\mu$ till $L=8$ have already been exhaustively found in \cite{fazeli2014scaling}. Consider $T_{5}$ (taken from \cite{bioglio2017minimum}) and their $E_{i,w}'s$ described in Table \ref{tableone}.

\begin{table}[!ht]
\centering
\caption{Polarization Behavior: Polynomial Coefficients $E_{i,w}$ of $T_{5}$}
\vspace{-0.1in}
\begin{tabular}{|c|c|c|c|c|c|c|}
\hline $i$ \textbackslash $w$ & 0 & 1 & 2 & 3 & 4 & 5  \\
\hline 0 & 0 & 3 & 9 & 10 & 5 & 1 \\
\hline 1 & 0 & 2 & 9 & 10 & 5 & 1 \\
\hline 2 & 0 & 0 & 2 & 8 & 5 & 1 \\
\hline 3 & 0 & 0 & 0 & 1 & 3 & 1 \\
\hline 4 & 0 & 0 & 0 & 1 & 2 & 1 \\
\hline
\end{tabular}
\label{tableone}
\end{table}

Using these values, we calculate $E_{i,w}$ using \eqref{eqnuse1} and \eqref{eqnuse2} for $T_{10}=T_{2}\otimes T_{5}$ and list it in Table \ref{tabletwo}. Once the values of polarization behavior are calculated for $L=10$, we calculate the scaling exponent of $T_{10}$ using \eqref{eqnuse3} and it comes out to be $\mu = 3.942$.
\begin{eqnarray*}
T_{10} & = &T_2 \otimes T_5, \text{where}\\
T_5 & =&\begin{bmatrix} 
 1 & 0 & 0 & 0 & 0 \\
0 & 1 & 0 & 0 & 0 \\
0 & 1 & 1 & 0 & 0 \\
1 & 1 & 0 & 1 & 0 \\
0 & 0 & 1 & 1 & 1 \\
 \end{bmatrix}
 \end{eqnarray*}

\begin{table}[ht!]
\centering
\caption{Polarization Behavior: Polynomial Coefficients $E_{i,w}$ of $T_{10}$}
\vspace{-0.1in}
\resizebox{\columnwidth}{!}{%
\begin{tabular}{|c|c|c|c|c|c|c|c|c|c|c|c|}
\hline $i$ \textbackslash $w$ & 0 & 1 & 2 & 3 & 4 & 5 & 6 & 7 & 8 & 9 & 10 \\
\hline 0 & 0 & 4 & 38 & 116 & 209 & 252 & 210 & 120 & 45 & 10 & 1\\
\hline 1 & 0 & 2 & 37 & 116 & 209 & 252 & 210 & 120 & 45 & 10 & 1\\
\hline 2 & 0 & 0 & 0 & 0 & 174 & 240 & 208 & 120 & 45 & 10 & 1\\
\hline 3 & 0 & 0 & 0 & 0 & 0 & 98 & 147 & 104 & 43 & 10 & 1\\
\hline 4 & 0 & 0 & 0 & 0 & 0 & 48 & 120 & 96 & 42 & 10 & 1\\
\hline 5 & 0 & 0 & 3 & 24 & 90 & 150 & 166 & 112 & 45 & 10 & 1\\
\hline 6 & 0 & 0 & 2 & 16 & 66 & 118 & 150 & 106 & 45 & 10 & 1\\
\hline 7 & 0 & 0 & 0 & 0 & 2 & 24 & 44 & 48 & 37 & 10 & 1\\
\hline 8 & 0 & 0 & 0 & 0 & 0 & 0 & 1 & 4 & 7 & 6 & 1\\
\hline 9 & 0 & 0 & 0 & 0 & 0 & 0 & 1 & 4 & 6 & 4 & 1\\
\hline
\end{tabular}
}
\label{tabletwo}
\end{table}

 Now, we introduce a heuristic approach to design a product kernel $T_{14} = T_2 \otimes T_7$ with good scaling exponent.

Let us take $8\times 8$ kernel giving best scaling exponent from \cite{fazeli2014scaling} and delete a row and a column in such a way that it gives us good polarizing behavior of the resulting kernel. As we know that the top and bottom channels in the initial kernel polarize to 1 and 0 respectively, we select a row whose polarization value is not close to both these values. In this case, we select the fourth row to delete. In the next step after deleting the row, we remove each column once and find scaling exponents to be $\{4.145, 4.110, 4.110, 4.129, 4.051, 3.984, 4.189\}$ obtained by deleting columns starting from $2^{nd}$ to $8^{th}$ (removing 1st column doesn't give a valid polarizing kernel). We find that good $\mu(T_7)$ is obtained when the seventh column is removed which almost coincides with the $\mu(T_7)$ found in \cite{fazeli2014scaling}. Therefore, we get $T_7$ to be

\begin{center}
$T_{7} = \begin{bmatrix} 
1 & 0 & 0 & 0 & 1 & 0 & 0\\
1 & 0 & 0 & 1 & 0 & 0 & 0\\
1 & 0 & 1 & 0 & 0 & 0 & 0\\
1 & 0 & 1 & 0 & 1 & 0 & 0\\
1 & 1 & 0 & 0 & 1 & 1 & 0\\
1 & 1 & 1 & 1 & 0 & 0 & 0\\
1 & 1 & 1 & 1 & 1 & 1 & 1 \end{bmatrix}$
\end{center}

\begin{table}[ht!]
\centering
\caption{Polarization Behavior: Polynomial Coefficients $E_{i,w}$ of $T_{7}$}
\vspace{-0.1in}
\begin{tabular}{|c|c|c|c|c|c|c|c|c|}
\hline $i$\textbackslash $w$ & 0 & 1 & 2 & 3 & 4 & 5 & 6 & 7 \\
\hline 0 & 0 & 4 & 18 & 34 & 35 & 21 & 7 & 1 \\
\hline 1 & 0 & 2 & 15 & 33 & 35 & 21 & 7 & 1 \\
\hline 2 & 0 & 0 & 9 & 31 & 35 & 21 & 7 & 1 \\
\hline 3 & 0 & 0 & 0 & 4 & 20 & 18 & 7 & 1 \\
\hline 4 & 0 & 0 & 0 & 2 & 10 & 15 & 7 & 1 \\
\hline 5 & 0 & 0 & 0 & 1 & 4 & 9 & 7 & 1 \\
\hline 6 & 0 & 0 & 0 & 0 & 0 & 0 & 0 & 1 \\
\hline
\end{tabular}
\label{tablethree}
\end{table}

\begin{table*}[t]
\centering
\caption{Polarization Behavior: Polynomial Coefficients $E_{i,w}$ of $T_{14}$}
\vspace{-0.1in}
\begin{tabular}{|c|c|c|c|c|c|c|c|c|c|c|c|c|c|c|c|}
\hline $i$ \textbackslash $w$ & 0 & 1 & 2 & 3 & 4 & 5 & 6 & 7 & 8 & 9 & 10 & 11 & 12 & 13 & 14 \\
\hline 0 & 0 & 0 & 64 & 336 & 984 & 1996 & 3002 & 3432 & 3003 & 2002 & 1001 & 364 & 91 & 14 & 1 \\
\hline 1 & 0 & 0 & 36 & 324 & 967 & 1990 & 3001 & 3432 & 3003 & 2002 & 1001 & 364 & 91 & 14 & 1 \\
\hline 2 & 0 & 0 & 0 & 252 & 933 & 1978 & 2999 & 3432 & 3003 & 2002 & 1001 & 364 & 91 & 14 & 1 \\
\hline 3 & 0 & 0 & 0 & 0 & 0 & 478 & 2021 & 2976 & 2856 & 1972 & 998 & 364 & 91 & 14 & 1 \\
\hline 4 & 0 & 0 & 0 & 0 & 0 & 0 & 1203 & 2560 & 2714 & 1942 & 995 & 364 & 91 & 14 & 1 \\
\hline 5 & 0 & 0 & 0 & 0 & 0 & 0 & 0 & 1840 & 2444 & 1882 & 989 & 364 & 91 & 14 & 1 \\
\hline 6 & 0 & 0 & 0 & 0 & 0 & 0 & 0 & 0 & 0 & 1642 & 525 & 280 & 84 & 14 & 1 \\
\hline 7 & 0 & 0 & 4 & 48 & 258 & 820 & 1714 & 2480 & 2547 & 1874 & 985 & 364 & 91 & 14 & 1 \\
\hline 8 & 0 & 0 & 2 & 24 & 135 & 470 & 1113 & 1848 & 2155 & 1746 & 969 & 364 & 91 & 14 & 1 \\
\hline 9 & 0 & 0 & 0 & 0 & 9 & 90 & 391 & 968 & 1499 & 1490 & 937 & 364 & 91 & 14 & 1 \\
\hline 10 & 0 & 0 & 0 & 0 & 0 & 0 & 4 & 32 & 116 & 248 & 322 & 232 & 79 & 14 & 1 \\
\hline 11 & 0 & 0 & 0 & 0 & 0 & 0 & 2 & 16 & 58 & 124 & 167 & 140 & 67 & 14 & 1 \\
\hline 12 & 0 & 0 & 0 & 0 & 0 & 0 & 1 & 8 & 28 & 56 & 73 & 68 & 43 & 14 & 1 \\
\hline 13 & 0 & 0 & 0 & 0 & 0 & 0 & 0 & 0 & 0 & 0 & 0 & 0 & 0 & 0 & 1 \\
\hline
\end{tabular}
\label{tablefour}
\end{table*}

The scaling exponent of $T_{7}$ comes out to be $\mu = 3.984$. We construct $T_{14} = T_{2}\otimes T_{7}$ and from the Table \ref{tablethree} values, we find polarization behavior of $T_{14}$ using \eqref{eqnuse1} and \eqref{eqnuse2} and are listed down in Table \ref{tablefour}. Scaling exponent for $T_{14}$ comes out to be $\mu = 3.485$.

\begin{figure}[!ht]
    \centering
    \includegraphics[scale=0.49]{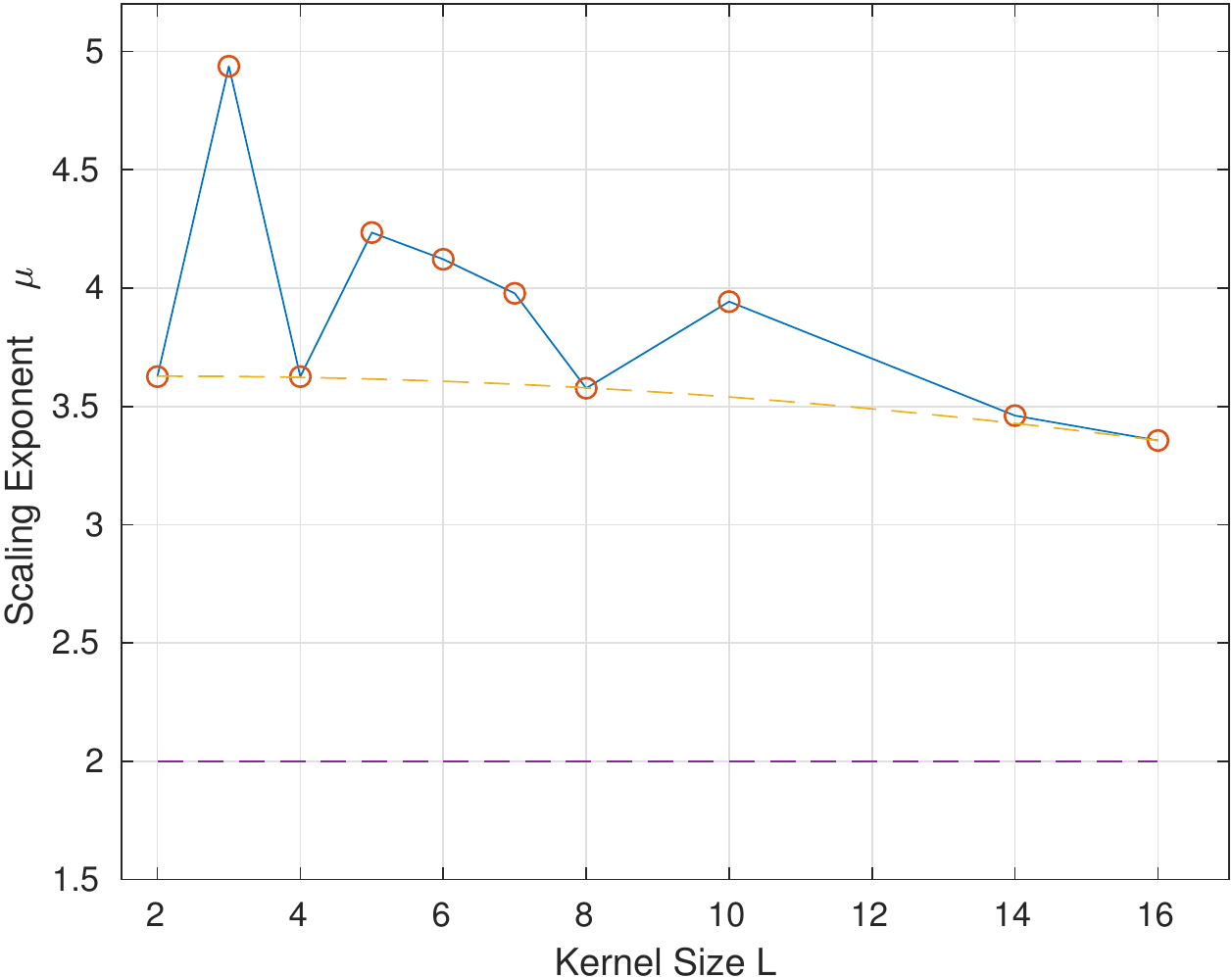}
    \caption{Scaling exponent of binary polarization kernel of size $L$}
    \label{fig:my_label2}
\end{figure}

A graph of the values of scaling exponent obtained for various $L$ for BEC channel $W$ keeping the probability of error constant is also plotted and is shown in Fig \ref{fig:my_label2}. The $\mu$ values for $2\leqslant L\leqslant 8$ have been taken from \cite{fazeli2014scaling} which are the best scaling exponent values. In this paper, we calculate $\mu$ for $L$ = \{10,14\} which comes out to be 3.942 and 3.485 respectively (these may not be the best scaling exponents for these lengths). We observe that for all powers of two on the x-axis, the scaling exponent gradually decreases as $L$ increases and approaches to 2 as $L$ tends to $\infty$. 

We also see from the plot that for all values of $L$ lying between two consecutive powers of two, values of scaling exponent are higher than the values at those two points. One of these values attain a local maxima which is always less than the maxima obtained for the previous consecutive powers of two.

\section{conclusion}
In this paper, we proved the property that any product kernel formed by taking kronecker product of self-dual component kernels is also self-dual. We also derived the partial distances of these product kernels in terms of the partial distances of the component kernel.

We proposed a method to find polarization behavior and scaling exponent of product kernels using polarization behavior of component kernels. We plot the behavior of scaling exponent with increasing kernel size. The scaling exponent for $L=10$ is calculated to be $\mu=3.942$ and for $L=14$ is $\mu=3.485$. 

%\section*{Acknowledgement}
%This work was supported partly by the Core Research Grant (CRG/2019/006637) from Science and Engineering Research Board (SERB) to V. Lalitha.

\bibliographystyle{IEEEtran}
	\bibliography{scaling_exponent}	

% Generated by IEEEtran.bst, version: 1.13 (2008/09/30)
\begin{thebibliography}{10}
\providecommand{\url}[1]{#1}
\csname url@samestyle\endcsname
\providecommand{\newblock}{\relax}
\providecommand{\bibinfo}[2]{#2}
\providecommand{\BIBentrySTDinterwordspacing}{\spaceskip=0pt\relax}
\providecommand{\BIBentryALTinterwordstretchfactor}{4}
\providecommand{\BIBentryALTinterwordspacing}{\spaceskip=\fontdimen2\font plus
\BIBentryALTinterwordstretchfactor\fontdimen3\font minus
  \fontdimen4\font\relax}
\providecommand{\BIBforeignlanguage}[2]{{%
\expandafter\ifx\csname l@#1\endcsname\relax
\typeout{** WARNING: IEEEtran.bst: No hyphenation pattern has been}%
\typeout{** loaded for the language `#1'. Using the pattern for}%
\typeout{** the default language instead.}%
\else
\language=\csname l@#1\endcsname
\fi
#2}}
\providecommand{\BIBdecl}{\relax}
\BIBdecl

\bibitem{arikan2009channel}
E.~Arikan, ``Channel polarization: A method for constructing capacity-achieving
  codes for symmetric binary-input memoryless channels, ieee t. inform. theory,
  55, 3051--3073,'' 2009.

\bibitem{tal2011list}
I.~Tal and A.~Vardy, ``List decoding of polar codes,'' in \emph{2011 IEEE
  International Symposium on Information Theory Proceedings}.\hskip 1em plus
  0.5em minus 0.4em\relax IEEE, 2011, pp. 1--5.

\bibitem{korada2010polar}
S.~B. Korada, E.~Sasoglu, and R.~Urbanke, ``Polar codes: Characterization of
  exponent, bounds, and constructions,'' \emph{IEEE Transactions on Information
  Theory}, vol.~56, no.~12, pp. 6253--6264, 2010.

\bibitem{mondelli2016unified}
M.~Mondelli, S.~H. Hassani, and R.~L. Urbanke, ``Unified scaling of polar
  codes: Error exponent, scaling exponent, moderate deviations, and error
  floors,'' \emph{IEEE Transactions on Information Theory}, vol.~62, no.~12,
  pp. 6698--6712, 2016.

\bibitem{fazeli2017binary}
A.~Fazeli, S.~H. Hassani, M.~Mondelli, and A.~Vardy, ``Binary linear codes with
  optimal scaling and quasi-linear complexity,'' \emph{arXiv: 1711.01339},
  2017.

\bibitem{yao2019explicit}
H.~Yao, A.~Fazeli, and A.~Vardy, ``Explicit polar codes with small scaling
  exponent,'' \emph{arXiv preprint arXiv:1901.08186}, 2019.

\bibitem{fazeli2014scaling}
A.~Fazeli and A.~Vardy, ``On the scaling exponent of binary polarization
  kernels,'' in \emph{2014 52nd Annual Allerton Conference on Communication,
  Control, and Computing (Allerton)}.\hskip 1em plus 0.5em minus 0.4em\relax
  IEEE, 2014, pp. 797--804.

\bibitem{benammar2017multi}
M.~Benammar, V.~Bioglio, F.~Gabry, and I.~Land, ``Multi-kernel polar codes:
  Proof of polarization and error exponents,'' in \emph{2017 IEEE Information
  Theory Workshop (ITW)}.\hskip 1em plus 0.5em minus 0.4em\relax IEEE, 2017,
  pp. 101--105.

\bibitem{hassani2014finite}
S.~H. Hassani, K.~Alishahi, and R.~L. Urbanke, ``Finite-length scaling for
  polar codes,'' \emph{IEEE Transactions on Information Theory}, vol.~60,
  no.~10, pp. 5875--5898, 2014.

\bibitem{bioglio2017minimum}
V.~Bioglio, F.~Gabry, I.~Land, and J.-C. Belfiore, ``Minimum-distance based
  construction of multi-kernel polar codes,'' in \emph{GLOBECOM 2017-2017 IEEE
  Global Communications Conference}.\hskip 1em plus 0.5em minus 0.4em\relax
  IEEE, 2017, pp. 1--6.

\end{thebibliography}
\end{document}